\documentclass[preprint,prc,aps,showpacs]{revtex4}
\usepackage{epsfig}
\begin{document}
\flushbottom

\title{Single and Double Photonuclear Excitations in Pb+Pb Collisions at $\sqrt{s_{NN}}$ = 2.76 TeV at the 
CERN Large Hadron Collider}
\author{{\O}ystein Djuvsland and Joakim Nystrand}
\affiliation{Department of Physics and Technology, 
\break University of Bergen, Bergen, Norway}

\begin{abstract}
Cross sections are calculated for single and double photon exchange in 
ultraperipheral Pb+Pb collisions at the LHC. The particle production 
is simulated with the DPMJET event generator. Large cross sections 
are found for particle production around mid-rapidity making these processes an 
important background to hadronic nuclear interactions at both the trigger and 
analysis levels. 
\\
\end{abstract}

\pacs{13.60.Le, 25.20.Lj, 25.75.-q} 

\maketitle

The strong electromagnetic fields present in high-energy nuclear collisions 
with impact parameters larger than the sum of the nuclear radii lead to 
large cross sections for a variety of photonuclear processes. In these 
ultraperipheral collisions, there are no geometrical overlaps 
between the colliding nuclei, so purely hadronic interactions are 
suppressed. Particle production in ultraperipheral collisions has been 
studied in Au+Au collisions at RHIC and also in heavy-ion interactions 
at lower energies. See Ref.~\cite{Bertulani:2005ru} for a review. These studies have 
so far been 
focused on the exclusive production of a single (vector) meson, e.g. 
$Au+Au \rightarrow Au+Au+V$ with $V = \rho^0$ or $J/\Psi$, or on two-photon
production of dilepton pairs. The nuclei have remained intact or have only 
been excited to low energies by the exchange of an additional soft photon. 

In this paper, we consider particle production in a general photonuclear
interaction, $\gamma + A \rightarrow X$ in ultraperipheral collisions 
between two lead nuclei at the CERN Large Hadron Collider (LHC). We do the 
calculations for the collision energy $\sqrt{s_{NN}} =$~2.76~TeV which will 
be available during the first two heavy-ion runs in 2010 and 2011/12. 
As will be 
discussed further below, the relevant photon energies for particle production 
around mid-rapidity are between $\sim 10$~GeV and $\sim 100$~TeV in the rest--frame of the 
target nucleus. We consider the cases when one or two photons are exchanged. 
In the former case, the photon-emitting nucleus remains intact, 
$A + A \rightarrow A + X$, while in the latter case both nuclei disintegrate. 

The particle production in the photonuclear interactions is modeled using 
the DPMJET Monte Carlo event generator version 3.0 \cite{Engel:1996yb,Roesler:1998wy}. 

We begin by discussing the photon flux associated with relativistic heavy ions 
and derive the relevant spectrum for single and double photon exchange. We then 
discuss particle production in these collisions and how the produced particles 
are distributed in phase space. We conclude by briefly discussing the experimental 
consequences of these processes in a typical high-energy collider experiment.

The electromagnetic field of a relativistic particle can be treated as 
an equivalent flux of photons. This is the Weizs{\"a}cker-Williams method. For 
collisions between relativistic nuclei, the photon spectrum should be 
calculated in impact parameter space \cite{Cahn:1990jk,Baur:1990fx}. 
In this way, interactions with 
nuclear overlap, where the strong interaction dominates, can easily be excluded. 
The difference between this approach and using a nuclear form factor for calculating 
the photon spectrum will be discussed below. 

The photon spectrum, $dn/dk$, of one ion in an ultraperipheral collision is given 
by an integral over the impact parameter, $b$,  
\begin{equation}
\frac{dn}{dk} = 
2 \pi \int_0^{\infty} \frac{d^3n}{dk db^2} (1 - P_{had}(b)) b db 
\label{single}
\end{equation}
The cut-off at small impact parameters $\sim 2R$ is implemented using a smooth 
function for the probability of not having any hadronic interaction $(1-P_{had}(b))$;  
$P_{had}(b)$ is calculated from a Glauber model. Since the photonuclear interaction 
probabilities are highest in grazing collisions with impact parameters $b \approx 2R$, 
it is important to take the hadronic interaction probability into account properly 
and not rely on a simple cutoff at $b=2R$ (see Fig.~\ref{probability} below). 
The doubly differential photon 
spectrum $d^3n/dk db^2$ is 
\begin{equation}
\frac{d^3n}{dk db^2} = \frac{Z^2 \alpha}{\pi^2} \frac{1}{k b^2} x^2 K_1^2(x)
\label{dndkdb}
\end{equation}
where $x = b k / \gamma$ and $\gamma >> 1$ is the Lorentz factor. 

The cross section for a photonuclear interaction with photons from a single
beam is then given by 
\begin{equation}
\sigma_{A+A \rightarrow A+X} =  
\int_{k_{min}}^{\infty} \frac{dn}{dk} \sigma_{\gamma A}(k) dk 
\label{total}
\end{equation}
The integral is cut off by the rapid decrease of the photon spectrum for $k > \gamma/R \approx$~120~TeV 
in the rest frame of the target nucleus. 

By inserting the expression for $dn/dk$ from Eq.~(\ref{single}) into Eq.~(\ref{total}) and changing the 
order of integration, the first order photonuclear interaction probability as a function 
of impact parameter can be obtained through differentiation:
\begin{equation}
P_1(b) = \frac{d \sigma}{db^2} = 
\int_{k_{min}}^{\infty} \frac{d^3n}{dk db^2} \sigma_{\gamma A}(k) dk
\end{equation}

The photonuclear interaction probability as function of impact parameter is shown in 
Fig.~\ref{probability} for three different values of $k_{min} =$~6~GeV, 1000~GeV, and 10~TeV. The 
minimum value, 6 GeV, is chosen as the lowest photon energy that can be handled by DPMJET. The 
effect of this minimum on the particle production in different rapidity intervals will be discussed 
further below. The photonuclear cross section, 
$\sigma_{\gamma A}(k)$, is the cross section for particle production calculated by DPMJET; 
it depends weakly on the photon energy, $k$, and increases monotonically from 8.1 to 12.9 mb over the 
relevant energy range (6~GeV -- 100~TeV). 

The total photon-proton cross section has not been measured above center-of-mass energies of 200 GeV, 
and there is thus some uncertainty in the energy dependence at the highest energies \cite{Block:2004ek}. 
The energy dependence of the photonuclear cross section should, however, be weaker because of shadowing, 
which implies that only the edges of the nuclei are affected by the increase. The contribution to the 
total cross section from photons with these high energies ($k >$~21~TeV) is furthermore rather low, so 
this uncertainty will not significantly affect the results presented here. 

The interaction probability has a maximum of $\approx$15\% for 
 $k_{min} =$~6~GeV in grazing collisions with impact parameter $\approx$16~fm. 
For $k_{min} =$~1000 GeV and 10 TeV, the corresponding maximum probabilities are 
6\% and 2\%, respectively. These high probabilities make exchange of 
multiple photons in the same event likely. For cases where the collision 
time $b/\gamma$ is much smaller than the excitation time, $1/k$, the sudden 
approximation applies, and the probability for multiple photon exchange 
factorizes \cite{Baur:2003ar}. This method has been used previously to calculate the total 
cross sections for correlated forward-backward Coulomb dissociation \cite{Baltz:1998ex}. 
The cross section for exchanging one photon 
from each nucleus is thus given by 
\begin{equation}
\sigma_{AA} = 2 \pi \int_{0}^{\infty} [P_1(b)]^2 (1 - P_{had}(b)) b db 
\label{totaldouble}
\end{equation}

The photon energy spectrum for double excitation is obtained by differentiating 
Eq.~(\ref{totaldouble}) with respect to the photon energies $k_1$ and $k_2$. The result
can be written 
\begin{equation}
\frac{d \sigma}{dk_1 dk_2} =  2 \pi \int_{0}^{\infty}  \frac{d^3n}{dk_1 db^2} \sigma(k_1)
\frac{d^3n}{dk_2 db^2} \sigma(k_2) (1 - P_{had}(b)) b db 
\label{double}
\end{equation}
It is worth noting that there is a positive correlation between the two photon energies. 
The photon spectrum from a single nucleus under the requirement that the other nucleus 
emits a photon is obtained by integrating over either $k_1$ or $k_2$. The result is 
\begin{equation}
\frac{d \sigma}{dk_1} = 2 \pi \int_{0}^{\infty}  \frac{d^3n}{dk_1 db^2} \sigma(k_1)
P_1(b) (1 - P_{had}(b)) b db 
\end{equation}
The spectrum is thus weighted toward a smaller impact parameter by the photonuclear 
interaction probability $P_1(b)$, resulting in a harder spectrum. This is in agreement 
with what was found for photonuclear vector meson production in coincidence with Coulomb 
break up \cite{Baltz:2002pp}. 

Although the interaction probabilities for these photon energies are fairly high, 
they are sufficiently smaller than unity for the first order probabilities to be used. 
Unitarity restoration \cite{Baur:2003ar} has been estimated to reduce the cross sections 
with at most a few percent.

A high energy photon may interact with a nucleon or nucleus in two different 
ways: it can appear as a bare photon, which interacts with a parton in the target, or 
it can first fluctuate to a $q \overline{q}$ pair, which then interacts with the 
target via the strong interaction, see e.g. Refs.~\cite{Drees:1988dk,Schuler:1993td}. 
The importance of the hadronic component of the photon at high-energy electron-proton colliders 
has been pointed out earlier \cite{Drees:1988dk}. Since the quantum numbers of the photon 
are $1^{--}$, it tends to fluctuate to a virtual vector meson (vector meson 
dominance). The bulk of the photonuclear cross section 
and the soft particle production can be understood from resolved vector meson interactions. 

The DPMJET \cite{Engel:1996yb,Roesler:1998wy} Monte Carlo event generator is based on the 
two-component dual parton model. 
The resolved part of the photon-hadron interaction is handled by the dual parton model through its 
implementation in PHOJET. The scaling from photon-hadron to photon-nucleus interactions, including 
shadowing, is treated within the framework of the Gribov-Glauber approximation. The high-$p_T$ 
partonic processes are calculated from lowest order perturbative QCD. The model has been tested against 
data from fixed target $\pi$-nucleus, p+nucleus, and $\mu$+nucleus experiments. It has been shown 
to reproduce the bulk particle production well \cite{Roesler:1998wy}. It can handle low or 
intermediate photon virtualities, which is fine in this case, since the photons from the 
Weizs{\"a}cker-Williams fields have very low virtualities $Q^2 \leq (1/R)^2 \approx$~10$^{-3}$~GeV$^2$. 
The fragmentation of the target nucleus is not considered in the current version of DPMJET, but 
knock-out protons from the target remnant are included. 

Samples of events have been generated with DPMJET with the photon spectrum described in the 
previous section and with different minimum photon energies. Six samples have been generated 
with 500,000 events per sample. Samples with single and double excitation have been generated 
with values of $k_{min} =$~6, 1000, and 10,000 GeV. The cross sections for these samples are listed in 
Table~\ref{cross_sections}. The table also includes the corresponding cross sections when it is 
required that there should be at least one charged particle at central rapidities. 

\begin{figure}
\epsfxsize=0.85\textwidth
\centerline{\epsffile{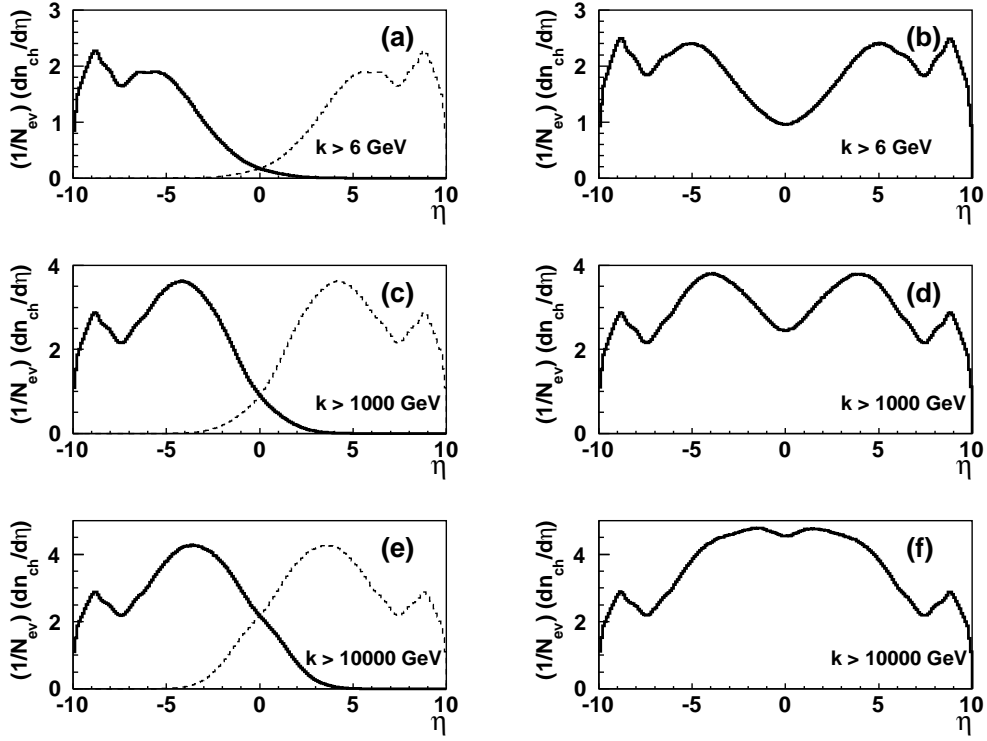}}
\caption[]{Pseudorapidity distributions of charged particles for different minimum photon energies for single (left) and double (right) excitations with different cut-off energies, $k_{min}$. For single excitations, the solid histograms are for photoproduction off the nucleus with negative rapidity and the dashed histogram corresponds to production off the nucleus with positive rapidity. The peaks around $\eta = \pm$~9 are from knock-out protons from the target nucleus.} 
\label{dndeta}
\end{figure}

\begin{figure}
\epsfxsize=0.85\textwidth
\centerline{\epsffile{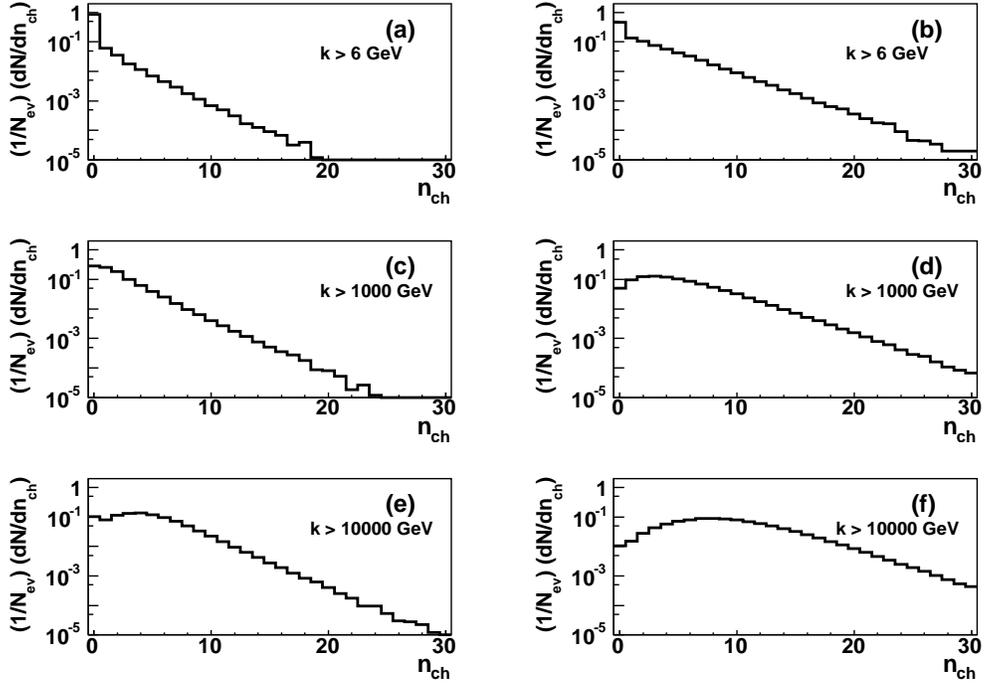}}
\caption[]{Multiplicity distributions of charged particles at mid-rapity $| \eta | <$~1 for single (left) and 
double (right) excitations with different cut-off energies, $k_{min}$.} 
\label{mult}
\end{figure}

The normalized charged particle pseudorapidity ($\eta = - \ln( \tan(\theta/2) )$) 
density distributions for the samples are shown in Figure~\ref{dndeta} (a) - (f). As expected, 
the particle production becomes more centered around mid-rapidity with increasing 
photon energy. The corresponding multiplicity distributions of charged particles 
with $| \eta | <$~1 are shown in Figure~\ref{mult} (a) - (f). 

The cross sections in Table~\ref{cross_sections} are huge. The total photoproduction 
cross section for $k > 6$~GeV is about 3 times larger than the total hadronic cross 
section. If one requires at least one photoproduced charged particle within 
$| \eta  | <$~1, the cross section is about 4~b or roughly 50\% of the total hadronic 
cross section. As can be 
inferred from Fig.~\ref{mult}, the probability for having larger multiplicities around 
mid--rapidity is also very high.  It is essential, especially for the higher values of 
$k_{min}$, to calculate the photon spectrum in the impact parameter space.  
If a form factor (a convolution of a hard sphere and Yukawa form factor \cite{Baltz:2002pp}) 
is used, the cross section for single excitation with $k_{min} =$~6~GeV has been found to be 
30\% higher than in Table I. For  $k_{min} =$~1000 and 10,000 GeV,
the differences are 73\% and 185\%, respectively. Using a nuclear form factor does not 
remove the contribution from events where the nuclei overlap. 

\begin{figure}
\epsfxsize=0.85\textwidth
\centerline{\epsffile{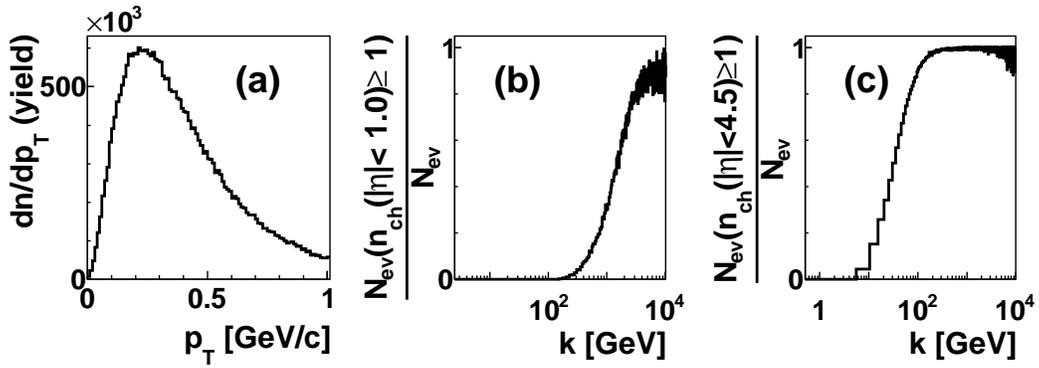}}
\caption[]{(a) Transverse momentum distribution of charged particles within $| \eta | <$~1. (b) Fraction of events with at least one charged particle within $| \eta | <$~1 as function of photon energy. (c) Same as (b) but for $| \eta | <$~4.5.} 
\label{ptratio}
\end{figure}

The transverse momenta of the photoproduced particles are typical for soft hadron-nucleus
interactions. The $p_T$ distribution at mid-rapidity for single production with 
$k_{min} =$~6 GeV is shown in Figure \ref{ptratio} (a). The distribution is essentially 
the same for all samples studied with a mean transverse momentum of $<p_T> \approx$~450~MeV/c. 

The large cross sections and the characteristics of these interactions should make them an important 
background to hadronic nuclear interactions at colliders. Single excitations, which have the largest cross 
sections, are characterized by a strong asymmetry around mid--rapidity event--by--event. They can 
thus to some extent be rejected by requiring the presence of particles on either side of mid--rapidity. 
The cross sections for double excitations are lower, but these events have particles produced over 
a wider range of rapidities. They may therefore be harder to separate from hadronic interactions 
with low or intermediate multiplicities. 

The photon spectrum of course extends to, in principle, arbitrarily low values of $k$. The 
lowest $k_{min} =$~6 GeV used here is therefore in that sense arbitrary. However, for 
particle production around mid-rapidity, photons with low energy do not contribute. This 
is illustrated in Figure \ref{ptratio} (b) and (c). Figure~\ref{ptratio} (b) shows the ratio of  
the number of events with at least one charged particle within $| \eta | < $1 to all events as 
a function of photon energy. The distribution goes to zero around $k \approx 150$~GeV, 
and photons with energy lower than this do thus not contribute to the particle production within 
the two most central units of pseudorapidity. Figure~\ref{ptratio} (c) shows the same distribution for 
events with at least one charged particle within $| \eta | < $4.5. The ratio goes to 
zero above the lowest $k_{min} =$~6 GeV used here. The photon energy range considered should 
thus give a complete description of photoproduction within the nine most central units of 
pseudorapidity, $| \eta | < $4.5. 

Unlike exclusive production of single mesons or dilepton pairs, inclusive photoproduction 
in ultraperipheral collisions has so far attracted rather limited interest. Some early 
studies were done for RHIC \cite{starnote}, mainly to find the 
background rates for exclusive production. The cross sections are about an order of magnitude 
lower at RHIC energies ($\sqrt{s_{NN}} =$~200~GeV) than at the LHC. Inclusive photoproduction of 
mesons in high energy heavy-ion interactions were also studied in Ref.~\cite{Chikin:2000my}. The results
are not directly comparable to the present work, because of a lower minimum photon energy, $k_{min}$, 
and because only multiple photons hitting the {\it same} nucleus were considered.

We have calculated the photon energy spectrum for single and double photonuclear excitations 
in Pb+Pb collisions at the LHC. Events have been generated with the DPMJET Monte Carlo 
according to these spectra. Large cross sections are found, also for particle production 
around mid-rapidity. 

The large cross sections and the non-zero probability for having charged particles 
produced with intermediate or high $p_T$ around mid-rapidity make these events 
an important background to peripheral and semi-central Pb+Pb collisions at the LHC. 
They have to be taken into account in order to correctly determine what 
fraction of the total hadronic cross section a given event selection corresponds to. 

If the photonuclear events can be clearly separated from the hadronic events, e.g. 
using the rapidity gap between the photon-emitting nucleus and the produced particles, much 
interesting physics could be extracted from them. The cross section to produce a 
$c \overline{c}$ pair through $\gamma$+gluon fusion is, for example, nearly 1~b at 
the full LHC energy \cite{Klein:2002wm}. If these events can be measured, they 
would provide valuable information on the nuclear parton distribution functions. 

We thank Stefan Roesler (CERN) for providing the code for the DPMJET Monte Carlo and
for help in getting it running. One of us (JN) would like to acknowledge 
useful discussions with Spencer Klein (Berkeley) and Anthony Baltz 
(Brookhaven) on multiple photon exchange in a single event.

\newpage 

\begin{table}
\caption[Table I]{Cross sections for particle production for single and double 
photonuclear excitations with different minimum photon energies, $k_{min}$.
The total cross sections and the cross sections for having at least one charged 
particle within $| \eta | <$~1 are shown. The cross section for single excitations 
take into account that both nuclei can act as photon--emitter and target.} 
\begin{tabular}{|c|c|c|c|c|}
\colrule
               & \multicolumn{2}{|c|}{single}             &  \multicolumn{2}{|c|}{double}               \\ \colrule
$k_{min}$ [GeV] & $\sigma$ [b] & $\sigma$ [b]              &  $\sigma$ [mb] &  $\sigma$ [mb]             \\ 
               & all          & $n_{ch}(|\eta|<1) \geq 1$  &  all           & $n_{ch}(|\eta|<1) \geq 1$   \\ 
   6           &  24.2        & 4.5                       &  240           &  130                       \\
  1000         &   4.9        & 3.5                       &  42            &  40                        \\
 10,000        &   0.90       & 0.81                      &  4.8           &  4.8                       \\ 
\colrule
\end{tabular}
\label{cross_sections}
\end{table}

\begin{figure}
\epsfxsize=0.6\textwidth
\centerline{\epsffile{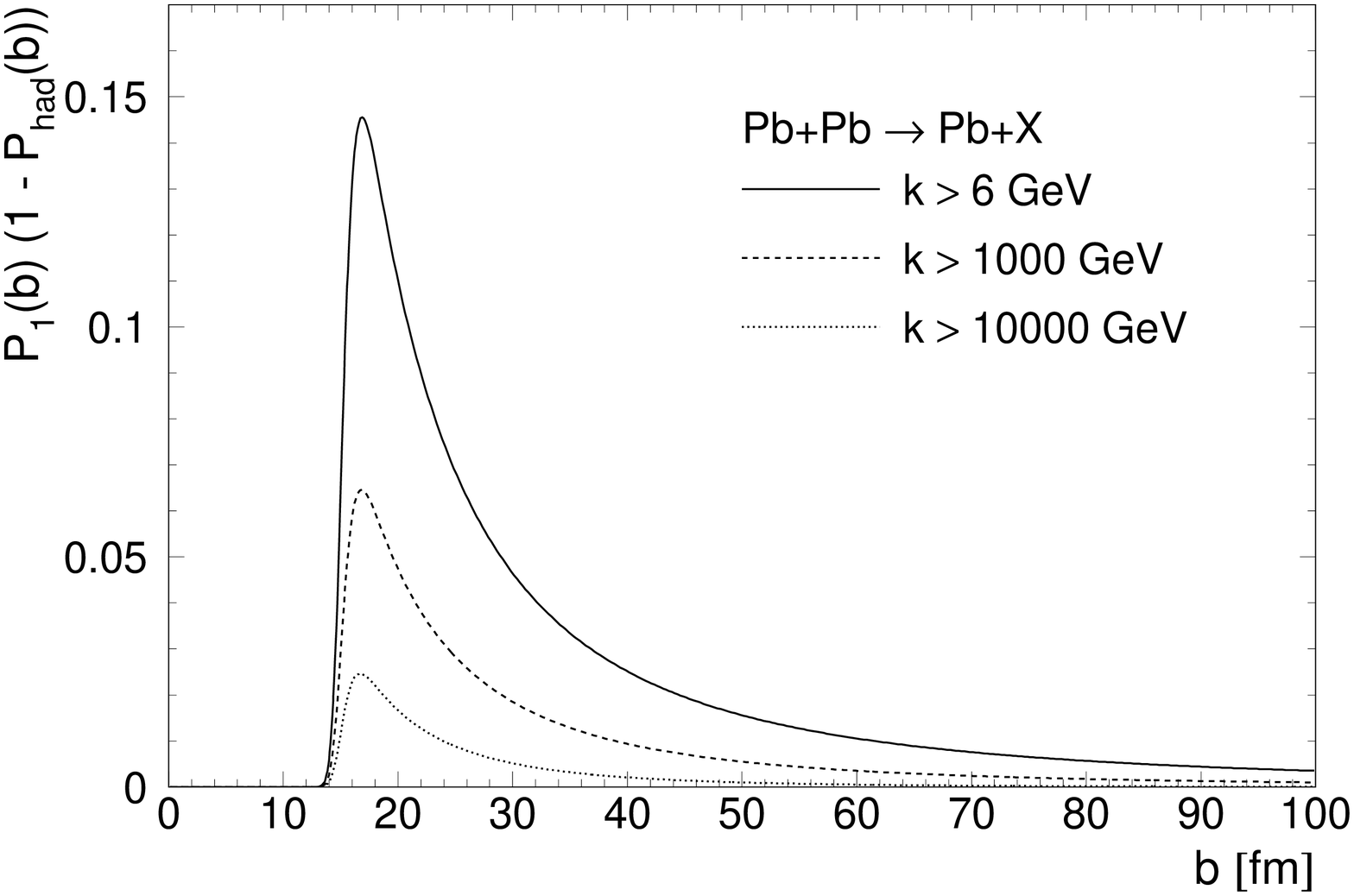}}
\caption[]{Probability for having a photonuclear interaction and no hadronic interaction as a function of 
impact parameter for three different minimum photon energies, $k_{min}$.}
\label{probability}
\end{figure}

\end{document}